\documentclass[sigconf, nonacm]{acmart}

\setcopyright{none}
\renewcommand\footnotetextcopyrightpermission[1]{}

\newcommand\vldbdoi{XX.XX/XXX.XX}

\newcommand\vldbavailabilityurl{https://github.com/LeiMa0324/KRONE_Demo_official}

\usepackage{amsthm}
\usepackage[normalem]{ulem} 
\usepackage[table]{xcolor} 

\usepackage{cases}
\usepackage{subcaption}
\makeatletter
\def\th@newremark{\th@remark\thm@headfont{\bfseries}}
\makeatletter
\theoremstyle{newremark}

\usepackage{tikz}

\usepackage[ruled]{algorithm}
\PassOptionsToPackage{noend}{algpseudocode}
\usepackage{algpseudocode}
\usepackage{amsmath}
\usepackage{bm}
\usepackage{color}
\usepackage{qtree}
\usepackage{graphicx}
\usepackage{multirow}
\usepackage{multicol}
\usepackage{url}
\usepackage{thmtools}
\usepackage{enumitem}

\usepackage{tabularx}
\usepackage{balance}    
\usepackage{booktabs}   
\usepackage{graphics}   
\usepackage{url}        
\usepackage{pifont}     

\usepackage{xcolor}      
\usepackage{xspace}
\usepackage{balance}

\usepackage[labelfont={bf,small},textfont={bf,small}]{caption}
\usepackage{pifont}
\usepackage{silence}
\renewcommand{\algorithmiccomment}[1]{\bgroup\hfill//~#1\egroup}

\newcommand{\eat}[1]{}

\definecolor{darkgreen}{rgb}{0.0, 0.5, 0.0}

\definecolor{ballblue}{rgb}{0.13, 0.67, 0.8}

\newcommand{\app}{\textsc{Krone}}
\newcommand{\apptree}{\textsc{Krone-Tree}}
\newcommand{\appviz}{\textsc{Krone-viz}}

\newcommand{\appseq}{\textit{\app{} Seq}}
\newcommand{\appseqs}{\textit{\app{} Seqs}}

\newcommand{\entityseqs}{\textit{\app{} E-seqs}}

\newcommand{\actionseqs}{\textit{\app{} A-seqs}}

\newcommand{\statusseqs}{\textit{\app{} S-seqs}}

\newcommand{\datamodel}{\textsc{\app{} Log Abstraction Model}}

\newcommand{\commentsymbol}{/*}
\algrenewcommand\algorithmiccomment[1]{\hfill \commentsymbol{} #1}

\renewcommand\footnotetextcopyrightpermission[1]{} 

\definecolor{myblue}{RGB}{135,206,250} 
\definecolor{lighterorange}{RGB}{255,225,180}  
 \definecolor{lightorange}{RGB}{255,200,130}
\definecolor{lightgray}{RGB}{240,240,240}  

\usepackage{booktabs}

\usepackage{tikz}
\usepackage{xcolor}

\definecolor{entityblue}{RGB}{218, 232, 252}
\definecolor{actiongreen}{RGB}{213,232,212}
\definecolor{statusorange}{RGB}{255, 230, 204}
\begin{document}

\title{Detect, Localize, and Explain: Interactive Hierarchical Log Anomaly Analytics with LLM Augmentation}



\author{Lei Ma}
\affiliation{
    \institution{Worcester Polytechnic Institute, USA}
    \streetaddress{}
    \city{}
    \state{}
    \country{}
}
\email{lma5@wpi.edu}

\author{Suhani Chaudhary}
\affiliation{
    \institution{UC Riverside, USA}
    \streetaddress{}
    \city{}
    \state{}
    \country{}
}
\email{schau062@ucr.edu}

\author{Ethan Shanbaum}
\affiliation{
\institution{Worcester Polytechnic Institute, USA}
  \streetaddress{}
  \city{}
  \state{}
  \country{}
}
\email{esshanbaum@wpi.edu}

\author{Athanasios Tassiadamis}
\affiliation{
    \institution{University of Nevada, Las Vegas, USA}
  \streetaddress{}
  \city{}
  \state{}
  \country{}
}
\email{tassia1@unlv.nevada.edu}

\author{Peter M. VanNostrand}
\orcid{0000-0002-0285-6019}
\affiliation{
  \institution{Worcester Polytechnic Institute, USA}
  \streetaddress{}
  \city{}
  \state{}
  \country{}
}
\email{pvannostrand@wpi.edu}

\author{Dennis M. Hofmann}
\affiliation{
  \institution{Worcester Polytechnic Institute, USA}
  \streetaddress{}
  \city{}
  \state{}
  \country{}
 }
\email{pvannostrand@wpi.edu}

\author{Haowen Xu}
\affiliation{
  \institution{Worcester Polytechnic Institute, USA}
  \streetaddress{}
  \city{}
  \state{}
  \country{}
 }
\email{hxu4@wpi.edu}

\author{Elke Rundensteiner}
\affiliation{
  \institution{Worcester Polytechnic Institute, USA}
  \streetaddress{}
  \city{}
  \state{}
  \country{}
 }
 \email{rundenst@wpi.edu}

\renewcommand{\shortauthors}{Ma, Chaudhary, and Shanbaum, et al.}
\begin{abstract}

Logs are ubiquitous in modern systems. Unfortunately, their unstructured nature in flat sequences 
limits understanding of execution behaviors, hindering effective anomaly diagnosis.
To address this, \app{} introduces a novel hierarchical log abstraction which transforms flat log sequences into semantically coherent units across entity, action, and status levels. Building on this abstraction, \app{} introduces a hierarchical orchestration framework that decomposes flat log sequences into hierarchical execution units and performs modular detection over them.
It executes and optimizes the 
modular detection tasks across levels, enabling precise anomaly detection, localization, and explanation with selective invocation of LLM-based reasoning.
In this work, we present \appviz{}, an interactive visualization system based on \app{}, which makes hierarchical log analysis 
interpretable 
and actionable for  
software engineers and system operators.   Demonstrated on the widely used HDFS benchmark dataset, \appviz{} supports:  
1) examine hierarchical decompositions of flat log sequences, 2) inspect detection results and abnormal segments identified by \app{} with LLM-generated explanations, 
, and 3) reuse-review-revise knowledge generated by LLMs with human in-the-loop guardrails. The code of \appviz{} is available at \url{https://github.com/LeiMa0324/KRONE_Demo_official}, and we deploy a live demo at \url{https://leima0324.github.io/KRONE_Demo_official}.

\end{abstract}

\maketitle




\section{Introduction}

\begin{figure}[t]
    \centering
    \includegraphics[trim={2.5mm 6mm 4mm 0}, clip, width=1.0\linewidth]{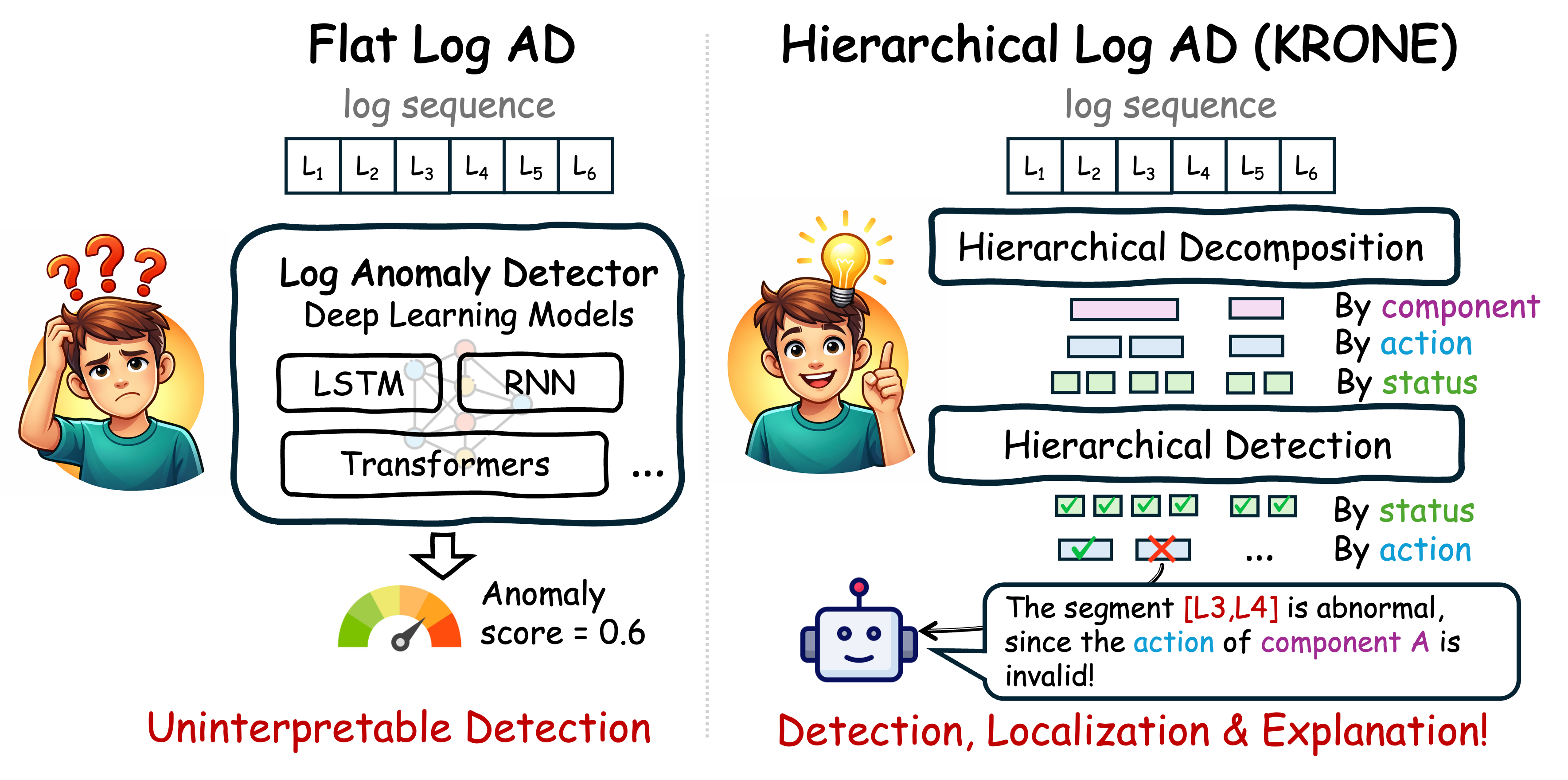}
    \vspace{-15pt}
    \caption{Flat log versus hierarchical log anomaly detection (\app{}).}
    \Description[]{}
    \label{fig:intro}
    \vspace{-15pt}
\end{figure}

\noindent\textbf{Motivation.} 
In modern data-intensive infrastructures, logs serve as a primary source for understanding system execution and diagnosing anomalies. Effective log analysis enables timely system monitoring, rapid issue resolution, and reliable operation at scale, making it essential for maintaining high-performance and dependable services.
%
%
However, as shown in Figure~\ref{fig:intro} left, most log anomaly detection methods \cite{deeplog,logbert, pluto} operate on \emph{flat log sequences} and produce a single anomaly score (label) for an entire log sequence. Further, they  have limited support for interpreting results and localizing anomalous segments. Recent LLM-based approaches \cite{Bridging_the_Gap,zhang2025xraglog} improve interpretability through natural language explanations, but continue to  operate on flat log sequences. Plus, they suffer from limits on context window sizes and prohibitively high computational costs for massive log data.

\vspace{0.3em}
\noindent\textbf{Our work: \app{} Framework.} \app{}~\cite{krone} (accepted at IEEE ICDE 2026) introduces the first hierarchical log abstraction that discovers and reconstructs execution semantics directly from logs. This abstraction enables  efficient, effective, and interpretable log anomaly detection.

Figure~\ref{fig:intro} compares \app{} with traditional flat log anomaly detection paradigms. The key insight of \app{} is that system executions follow hierarchical structures with nested components and operations. Accordingly, \app{} defines a three-level semantic hierarchy: Entities correspond to system components, Actions capture operations, and Statuses reflect the outcome of each action. Based on this hierarchy, \app{} decomposes log sequences into atomic and reusable execution segments and formulates anomaly detection as modular tasks across semantic levels, enabling precise localization of anomalies within execution contexts.

To leverage foundation models while minimizing cost, \app{} adopts a lightweight pattern-matching detector for fast filtering known normals and selectively invokes LLM-based reasoning for uncertain cases. \app{}  maintains a hierarchical knowledge base that reuses the modular detection results to support  knowledge-empowered analysis with further reduction in 
LLM calls. Experiments on real-world public benchmarks demonstrate the comprehensive improvement of \app{}, on accuracy (F1-score 82.76\% $\rightarrow$ 92.83\% over prior methods), data-efficiency (data space 117.3$\times\downarrow$), and LLM resource-efficiency (1.1\%–3.3\% of the test data size). We refer to \app{}~\cite{krone} for indepth
evaluation study.

\vspace{0.3em}
\noindent\textbf{\appviz{} Demonstration.} \app{} integrates anomaly detection, localization, and explanation with efficient LLM usage, reducing manual log analysis effort. To showcase its capabilities, we present \appviz{}, an interactive platform on the real-world HDFS log dataset. \appviz{} improves transparency by enabling engineers to explore hierarchical execution, inspect intermediate steps, and review or correct detection results, supporting human-in-the-loop analysis while mitigating LLM hallucination risks.
Specifically, our
demonstration
providing the audience
first-hand experience 
with 
 the following 
 \appviz{}
 innovations:


 \begin{figure}[t]
    \centering 
    \includegraphics[trim={3.3cm 1.6cm 0.2cm 1.9cm},clip, width=1.0\linewidth]{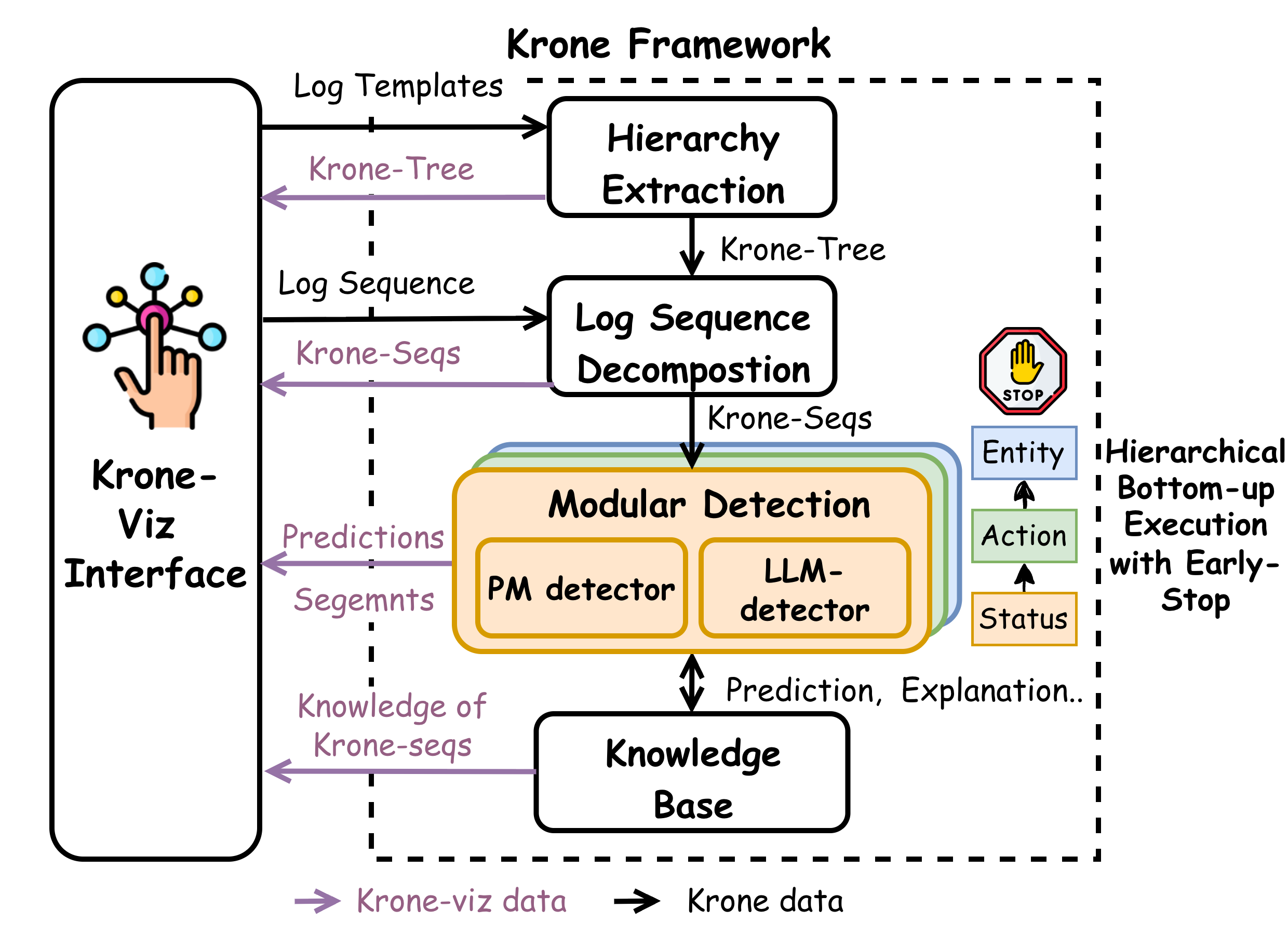}
    \vspace{-13pt}
    \Description[]{}
    \caption{
            \app{} framework overview.
            \label{fig:arch}
        }
   \vspace{-20pt}
\end{figure}

\begin{enumerate}[topsep=0mm, itemsep=0mm, wide]
    \item\textit{Hierarchical Log Abstraction.} 
    Through \app{}, users can see how \app{} transforms flat log sequences into an interpretable hierarchy of system execution, enabling decomposition into structured execution segments.
Mirroring expert reasoning, this enables users to understand the execution via semantic-segmented log sequences, without requiring domain knowledge.

    \item\textit{Hierarchical and Modular Anomaly Detection.}
\app{} performs modular anomaly detection over execution units and hierarchically composes the results, with LLM-generated explanations. Users can explore the hierarchical context of a long log sequence and rapidly identify the abnormal segments with explanations.


    \item\textit{Knowledge Accumulation and Revision.} We showcase how \app{} leverages atomic execution units shared across log sequences to build a hierarchical knowledge base for LLM result reuse. Users can inspect the detected anomalies with LLM-generated explanations, explore accumulated knowledge across sequences, and revise stored knowledge to guide subsequent analysis. 
    

\end{enumerate}

\section{The \app{} Framework}

We briefly introduce the \app{} framework and its core innovations.




\subsection{\app{} Framework Overview}
Figure~\ref{fig:arch} presents an overview of the \app{} framework, which consists of four modules. (1) \textit{Hierarchy Extraction} uses LLMs to extract the entity–action–status semantics from log templates\footnote{Following standard practice, raw log messages are first parsed into templates. A log sequence is then represented as a sequence of templates. Pre-processing details can be found in \cite{krone}.} to construct the semantic hierarchy, namely \apptree{}. (2) Using \apptree{}, \textit{Log Sequence Decomposition} transforms flat logs into hierarchical execution units (\appseqs{}). (3) \textit{Modular Detection} performs anomaly detection on each \appseq{} using lightweight pattern matching with selective LLM reasoning, orchestrated in a bottom-up manner with early stopping. (4) \textit{Knowledge Base} stores detection results and explanations for reuse, reducing redundant LLM calls. \appviz{} visualizes the intermediate outputs of each module, which is introduced in Sec. \ref{sec:demo}. 


\subsection{Core Innovations of \app{}}

\app{} is built on three core innovations: (1) the \datamodel{}, (2) the Hierarchical and Modular Anomaly Detection, and (3) the Knowledge Base for Amortized Detection. 
\begin{figure*}[t]
    \centering
    \includegraphics[trim={0 12mm 0 9.5mm},clip, width=1.0\linewidth]{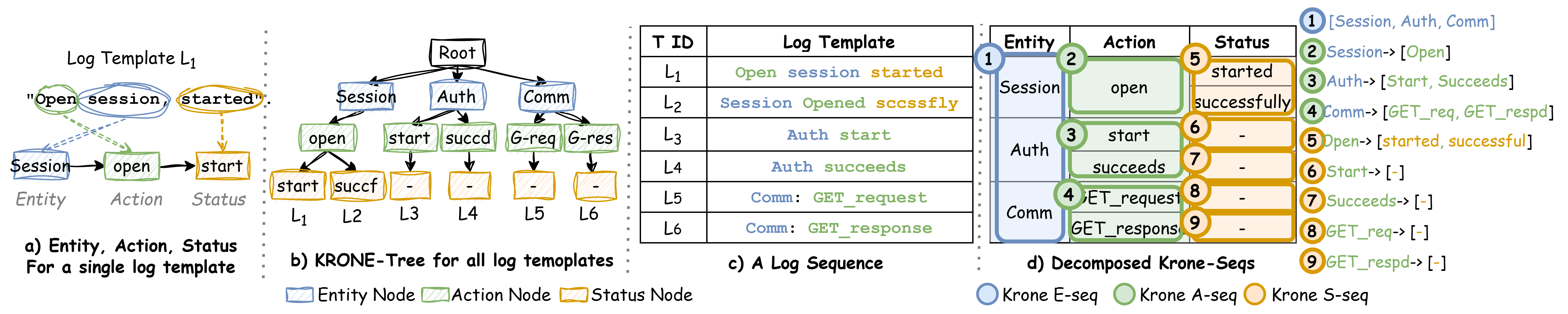}
    \vspace{-14pt}
    \Description[]{}
    \caption{\app{} Log Abstraction Model a) The entity, action, and status of a log template, b) The \apptree{} which encodes the semantic hierarchy from all log templates, c) An example log sequence of templates, and d) The \appseqs{} decomposed from the log sequence via \apptree{}, in the form of "<parent>$\rightarrow$ <node sequence>". Figure is adapted from the full paper \cite{krone}. }
    \label{fig:data_model}
\end{figure*}

\vspace{0.3em}
\noindent\textbf{Innovation 1: \datamodel{}. }
The key innovation of \app{} is a hierarchical abstraction for log modeling that can be instantiated as a dataset-specific semantic hierarchy (see Figure~\ref{fig:data_model}). This abstraction is based on the observation that log messages follow a consistent semantic pattern: a {\it status} describes an {\it action} performed on an {\it entity}. For example, Figure~\ref{fig:data_model}(a) shows the template “Open session started”, where "$Session$", "$Open$", and "$Started$" correspond to the entity, action, and status, respectively. Following this abstraction, \datamodel{} defines a semantic hierarchy, namely \apptree{}, over all log templates (Figure~\ref{fig:data_model}(b)), capturing entities, actions, and statuses along with their relationships. In \app{}, this hierarchy is extracted via an LLM-based NER formulation, where LLMs identify entity, action, and status from each template and construct the \apptree{}. Each leaf status node maps to a single template, while actions and entities group semantically related templates.

One key benefit of the \apptree{} is that it enables decomposition of long log sequences into semantically coherent units at multiple granularities, supporting a divide-and-conquer approach to anomaly detection. Given the \apptree{} and input log sequences, \app{} decomposes each sequence into \appseqs{}—contiguous segments represented as node sequences in the hierarchy—across entity, action, and status levels. As illustrated in Figure~\ref{fig:data_model}(c)--(d), \entityseqs{}, \actionseqs{}, and \statusseqs{} capture transitions at the three levels with the corresponding segment (as corresponding template rows), enabling hierarchical learning of normal patterns and modular anomaly detection within smaller scopes.

\vspace{0.3em}
\noindent\textbf{Innovation 2: Hierarchical and Modular Anomaly Detection.}
\app{} introduces a hierarchical, modular formulation that decomposes anomaly detection into semantically scoped tasks over \appseqs{}, enabling precise localization and scalability. Specifically, each \appseq{} is evaluated against its normal counterparts within a well-defined semantic scope determined by the \apptree{}. 
For modular detection, \app{} first applies a lightweight pattern-matching (PM) detector to efficiently filter out known normal \appseqs{}, and then augments detection by selectively invoking LLM-based reasoning to examine the remaining unknown \appseqs{}. Using $k$ normal \appseqs{} as in-context examples, the hierarchy-defined scope enables context-based verification and explanation by grounding a test \appseq{} in normal behavioral patterns, such as typical action transitions within an entity. 

The orchestration of modular detection in \app{} is guided by two properties of \appseqs{}: higher-level \appseqs{} are compositions of lower-level ones and incur higher LLM cost due to longer-range dependencies. This motivates a hierarchical, bottom-up strategy that examines \statusseqs{} first and proceeds upward only if no anomaly is detected. Since anomalies propagate through composition, \app{} employs early stopping upon detecting the first anomaly to avoid unncessary computation cost.



\vspace{0.3em}
\noindent\textbf{Innovation 3: Knowledge Base for Amortized Detection.}
Our empirical evaluation observes that \appseqs{} are atomic execution units that frequently recur across log sequences, often appearing in different compositions (permutations). This property makes their knowledge inherently reusable: once an \appseq{} is evaluated, its result can be reused across sequences, avoiding redundant LLM invocations. 
Following the same LLM-cost optimization principles, \app{} maintains a hierarchical knowledge base that stores LLM-generated artifacts for \appseqs{}, including detection outcomes and explanations. During modular detection, \app{} first queries this knowledge base for a given test \appseq{}, invoking the LLM only when no matching entry is found. As stored \appseq{} knowledge is semantically grounded and shared across sequences, this design enables substantial amortized cost reduction over time while improving interpretability.

Beyond caching, this knowledge base serves as a foundation for continuous refinement: newly detected patterns and human-validated corrections can be incorporated to improve future detection and explanation quality.  This design elevates LLM usage from per-instance inference to a knowledge-driven process that improves efficiency, interpretability, and adaptability over time.

\section{Demonstration of \appviz{}}
\label{sec:demo}

\begin{figure*}[t]
    \centering 
    \includegraphics[trim={0.1cm, 0cm, 0.65cm, 0cm}, clip, width=0.9\linewidth]{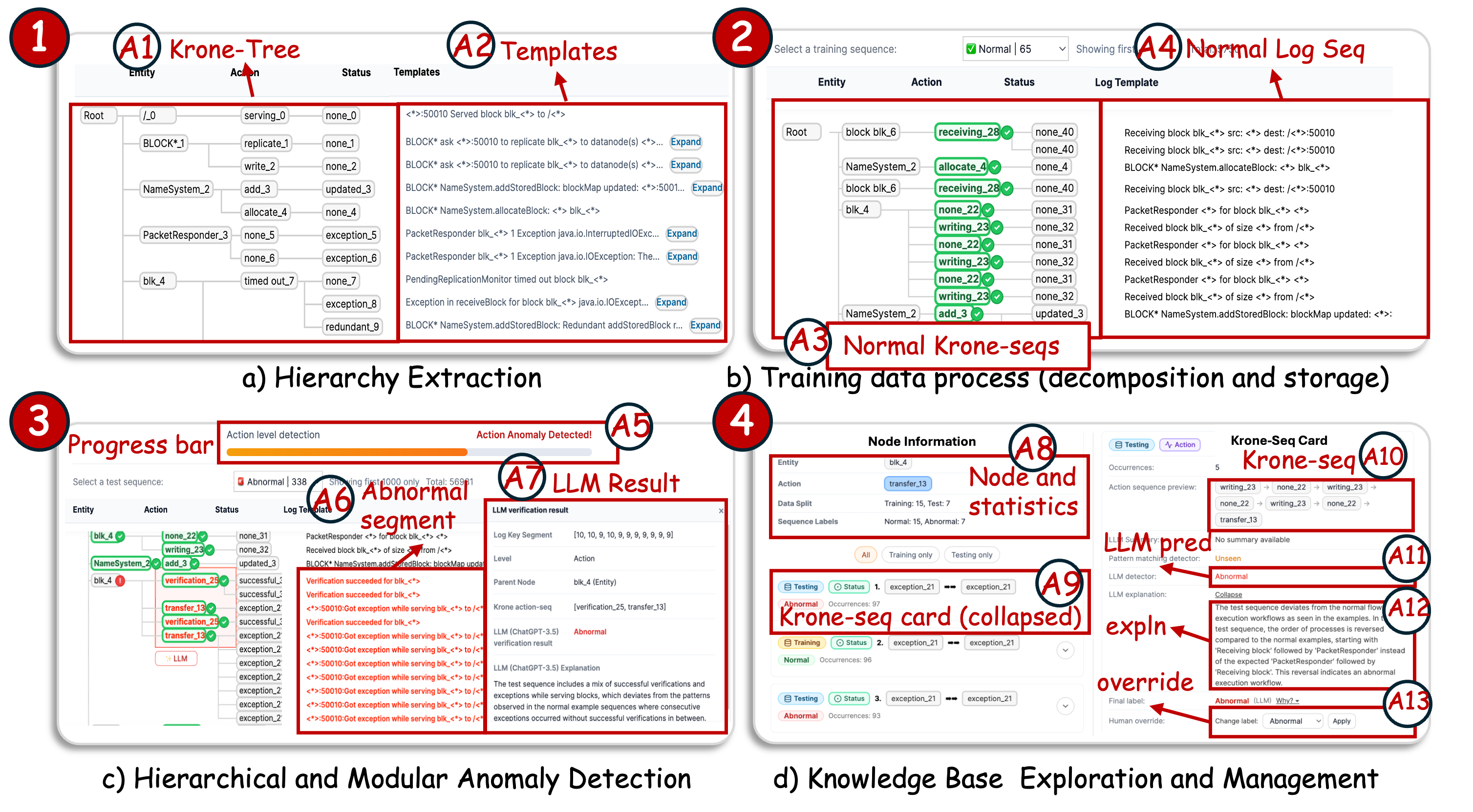}
    \vspace{-10pt}
    \caption{\appviz{} Interactive Platform. }
    \label{fig:demo}
    \Description[]{}
\end{figure*}



In this demonstration, the audience will experience the power of \app{} through \appviz{}, an interactive system with interlinked, customized views (Figure~\ref{fig:demo}). We showcase \app{}’s core innovations and workflows on the real-world HDFS public benchmark, with representative visualizations shown in Figure~\ref{fig:demo}. \appviz{} enables effective human-centered analysis by making \app{}’s hierarchical decomposition, anomaly localization, LLM-based explanations, and accumulated knowledge transparent and actionable. The full interactive experience is available on our demo site\footnote{\url{https://leima0324.github.io/KRONE_Demo_official/}}.






Through \appviz{}, the audience actively explores each core step of \app{} in the following four stages:

\begin{itemize}[wide, topsep=0mm]

\item\textit{Stage 1: Hierarchy Extraction.}
The audience begins by loading log templates into \appviz{}, which invokes an LLM to extract entities, actions, and statuses and construct the \apptree{} (Figure~\ref{fig:demo}(a)). The interface displays the resulting hierarchy along with tree statistics, allowing the audience to explore relationships between the \apptree{} (A1) and log templates (A2). By hovering over nodes, the audience can see associated templates, making the semantic structure of logs explicit. This stage highlights Innovation~1 by showing how \app{} transforms flat logs into an interpretable hierarchical abstraction.

\item \textit{Stage 2: Training Data Processing.}
The audience selects a normal log sequence and observes how \appviz{} decomposes it into hierarchical \appseqs{} (A3) aligned with the original sequence (A4) (Figure~\ref{fig:demo}(b)). As the decomposition proceeds, intermediate \statusseqs{} are stored and marked as completed at their parent nodes, illustrating the construction of ground-truth execution units. Through this interactive process, the audience learns how normal execution patterns are captured and stored in the knowledge base. \appviz{} also provides a batch mode for one-click processing of all training sequences.

\item\textit{Stage 3: Hierarchical and Modular Anomaly Detection.}
The audience selects a test log sequence and triggers anomaly detection (Figure~\ref{fig:demo}(c)). \appviz{} decomposes the sequence into test \appseqs{} and performs modular detection in a bottom-up manner from status to entity levels, with a progress bar (A5) indicating the current stage. For each \appseq{}, a pattern-matching detector first searches the knowledge base. If no match is found, the system flags the \appseq{} as potentially anomalous and highlights the corresponding segment (A6). The audience can then invoke the LLM to verify and explain the anomaly (A7). If confirmed, \app{} stops further analysis and reports the detected anomaly. This stage demonstrates Innovation~2 by enabling precise, cost-efficient anomaly localization and explanation.

\item\textit{Stage 4: Knowledge Base Exploration and Management.}
The audience explores and manages accumulated knowledge through the knowledge base interface (Figure~\ref{fig:demo}(d)). By interacting with nodes in the \apptree{}, the audience can browse summary statistics (A8) of each node, including its ancestors and the statistics of its \appseqs{}. The stored \appseqs{} are listed as collapsed cards (A9), including sources, semantic levels, predictions, and frequencies. Expanding a card reveals detailed results (A10–A13), such as pattern-matching outcomes, LLM predictions (A11), and explanations (A12). The audience can also revise detection results via an override mechanism (A13), updating the knowledge base for future reuse. This stage highlights Innovation~3 by demonstrating how reusable knowledge reduces LLM cost and supports continuous refinement.

\end{itemize}

\balance
\section{Conclusion}
We present \appviz{}, an interactive system for hierarchical log anomaly analytics based on \app{}. By exposing hierarchical structures, modular detection, and reusable knowledge, \appviz{} enables transparent anomaly detection, localization, and explanation. This demonstrates how \app{} achieves interpretable and scalable log analysis with efficient cost-aware LLM integration.

\section{Acknowledgement}
This work is supported in part by NSF through NRT-HDR-2021871, IIS-1910880, CSSI-2103832, and CNS-2349370. 
We thank our collaborators at ByteDance and Prof. Lei Cao (University of Arizona) for their valuable feedback.


\bibliographystyle{ACM-Reference-Format}
\bibliography{main}
\end{document}